# Structural Evolution of One-dimensional Spin Ladder Compounds $Sr_{14-x}Ca_xCu_{24}O_{41}$ with Ca doping and Related Hole Redistribution Evidence

Guochu Deng[a,b*], Vladimir Pomjakushin[c], Václav Petříček[d], Ekaterina Pomjakushina[a], Michel Kenzelmann[a], Kazimierz Conder[a]

[a] Laboratory for Developments and Methods, Paul Scherrer Institute, CH-5232 Villigen, Switzerland

[b] Bragg Institute, Australia Nuclear Science and Technology Organization, New Illawarra Rd, Lucas Height, NSW 2233

[c] Laboratory for Neutron Scattering, Paul Scherrer Institute, CH-5232 Villigen, Switzerland

[d] Institute of Solid State Physics, Czech Academy of Sciences, Cukrovarnická 10, 162 00 Phaha 6, Czech Republic

Abstract

Incommensurate crystal structures of spin ladder series $Sr_{14-x}Ca_xCu_{24}O_{41}$ (x=3, 7, 11, 12.2) were characterized by powder neutron scattering method and refined using the superspace group Xmmm(00γ)ss0 (equivalent to superspace group Fmmm(0,0,1+γ)ss0); X stands for non-standard centering (0,0,0,0), (0,½,½,½), (½,½,0,0), (½,0,½,½)) with a modulated structure model. The Ca doping effects on the lattice parameters, atomic displacement, Cu-O distances, Cu-O bond angles and Cu bond valence sum were characterized. The refined results show that the $CuO_4$ planar units in both chain and ladder sublattices become closer to square shape with an increase of Ca doping. The Cu bond valence sum calculation provided new evidence for the charge transfer from the chains to ladders (approximately 0.16 holes per Cu from *x*=0 to 12.2). The charge transfer was attributed to two different mechanisms: (a) the Cu-O bond distance shrinkage on the ladder; (b) increase of the interaction between two sublattices, resulting in Cu-O bonding between the chains and ladders. The low temperature structural refinement resulted in the similar conclusion, with a slight charge backflow to the chains.

---

[*] Corresponding author: guochu.deng@ansto.gov.au or guochu.deng.cn@gmail.com

I. INTRODUCTION

According to theoretical predictions, the S=1/2 one-dimensional antiferromagnetic Heisenberg magnetic structures, e.g. chains and ladders, exhibit a spin liquid state, which can be gapped or gapless depending on the spin configuration of the system. It was predicted that dimmerized chains and even-leg ladders should show magnetic excitations with specific energy gaps.[1] Furthermore, the hole-doped even-leg ladders were believed to become superconducting due to the hole pairing by coupling through the rungs, which is a completely different pairing mechanism from that proposed for cuprate superconductors.[1] Therefore, the theoretical work mentioned above raised extensive research attention in one-dimensional magnetic systems, such as spin ladder systems, of which $Sr_{14-x}Ca_xCu_{24}O_{41}$ is a particularly interesting case.[2]

$Sr_{14-x}Ca_xCu_{24}O_{41}$ is a self hole-doped one dimensional magnet, containing both edge-sharing $CuO_2$ chains and two-leg $Cu_2O_3$ ladders which stack along *b* axis separated by Sr/Ca layers.[2] The undoped parent phase $Sr_{14}Cu_{24}O_{41}$ is a charge density wave insulator while in the highly Ca-doped sample superconducting state can be induced by applying external pressure.[3] This is the only superconductor with a spin ladder structure, and so a natural question is if the mechanism of superconductivity in this material is the one predicted by Dagotto et al.[1] Till now no clear answer to this question was found even though many experimental efforts have been made, including inelastic neutron scattering,[4] nuclear magnetic resonance,[5] optical conductivity,[6] and many others. One complicating factor for the microscopic understanding of this material is the complex incommensurate structure.[7,8] Both Ca doping and external hydrostatic pressure are believed to be essential for the superconductivity in this system,[9] and the latter is believed to be an extension of the chemical pressure introduced by Ca doping. From this point view, study of the structural evolution of Ca doping in $Sr_{14-x}Ca_xCu_{24}O_{41}$ may shed light on the superconductivity mechanism in this series of compounds. The influence of Ca doping on magnetic and electrical properties has been widely investigated. However, the Ca doping effects on the crystal structure have not been studied in detail.

$Sr_{14-x}Ca_xCu_{24}O_{41}$ has an incommensurate crystal structure, which is composed of chain and ladder sublattices modulated along *c* axis. Therefore, it has to be described by a four dimensional (3+1) superspace formalism.[10] The early research work made by E. McCarron *et al.*[11] revealed that the crystal structure is composed of two subsystems. However, these authors did not solve the incommensurate structure of this material because they merely took into account the main reflections for the refinement and ignored the interaction between the two sublattices. In addition, no four dimensional superspace group formalism was used. After their pioneering work, several research groups proposed improvements in the refinement models, including four dimensional superspace group symmetry.[7, 8, 12, 13] However, all these models were based on different selections of the superspace groups for samples with different Ca doping levels, being subgroups of the highest superspace group $Fmmm(0,0,\gamma)ss0$.[14] For example, K. Ukei *et al.*[12] reported their structure model for the undoped parent compound $(Sr_2Cu_2O_3)_{0.7}CuO_2$ (i.e. $Sr_{14}Cu_{24}O_{41}$) with an acentric superspace group $A2_1ma(01\gamma)0s0$ for both chain and ladder subsystems. Jensen *et al.*[7, 8] proposed two different structure models, a centrosymmetric superspace group $Amma(0,0,1+\gamma)ss0$ for slightly Bi doped compound $(Sr_{1.92}Bi_{0.08}Cu_2O_3)_{0.7}CuO_2$ (i.e. $Sr_{13.44}Bi_{0.56}Cu_{24}O_{41}$), and an acentric space group $F222(0,0,1+\gamma)000$ for the intermediately Ca-doped compound $(Sr_{0.92}Bi_{0.12}Ca_{0.96}Cu_2O_3)_{0.7}CuO_2$ (i.e. $Sr_{6.44}Bi_{0.84}Ca_{6.72}Cu_{24}O_{41}$).[7, 8] A superspace group $F222(11\gamma)$: $F222(11\gamma)$ was also used by T. Ohta *et al.*[3] to refine neutron powder diffraction data for highly Ca doped sample. More recently, Y. Gotoh *et al.*[13] used superspace group $Amma(01\gamma)ss0$ :$Acca(01\gamma)$ to refine the crystal structure of the undoped sample and proposed a hole ordering model for the chain sublattice on the basis of the refined result. However, all these works were mainly focused on the structure of one sample with a special composition. Obviously, these results are not sufficient to systematically analyze the Ca doping effect on the crystal structure. Since the Ca doping is indispensible for this compound to transform into the superconducting state, it is highly important to clarify the Ca doping effect on the crystal structure.

As mentioned above, the $Sr_{14-x}Ca_xCu_{24}O_{41}$ system is a self hole-doped system, containing around 6 holes in one chemical formula unit, which was believed to be the critical feature contributing to many phenomena, e.g. dimerized singlet ground state, superconductivity, etc.[2] Since $Sr_{14-x}Ca_xCu_{24}O_{41}$ is constructed by two sublattice systems, the nonhomogeneous hole distribution is expected on these two subsystems because they have nonequivalent Cu-O bonding geometries. V.

Kato *et al.*[15] and Y. Gotoh *et al.*[13] estimated Cu valence on the chain sites being 2.38 and 2.55 in the undoped parent phase, respectively. While the Cu valence on the ladder sites was kept almost undoped. Even though isovalent Ca substitution on Sr sites is believed not to change the whole doped hole number in the formula unit, the holes were believed to transfer from the chain sublattice to the ladder sublattice with the increase of Ca doping as evidenced by optical conductivity measurement. However, how many holes were transferred by such a doping is still an open question. Obviously, this is a very important issue for many theoretical models and interpretations of the phenomena observed in this system, e.g., hole crystallization, chain spin dynamics, etc. Therefore, it is essential to find out a possible structural evidence for the charge transfer between the two sublattices.

In this study, we investigated structural evolution of the $Sr_{14-x}Ca_xCu_{24}O_{41}$ in function of Ca doping using neutron powder diffraction. The results were analyzed by means of Rietveld refinement using the (3+1) superspace formalism. Our refined results indicate that the crystal structure of the system shrinks and becomes closer to orthogonal with increasing Ca doping. By substituting more Ca on Sr sites, the $CuO_4$ rectangles in the chains reduce their length/width ratio; all the O-Cu-O angles become closer to 90°; the rungs become straighter and the rung lengths on the ladders reduce. Furthermore, using the bond valence sum (BVS) method, we quantitatively estimate the Cu valence on the chains and ladders for samples with various Ca contents, providing structural evidence for charge transfer from chains to ladders, confirming the previous optical measurement.[6]

II. STRUCTURAL MODEL

The crystal structure of $Sr_{14-x}Ca_xCu_{24}O_{41}$ series is composed of two sublattices, $CuO_2$ chain, and $(Sr_{1-x}Ca_x)_2Cu_2O_3$ ladder, which have common *a* and *b* axis but different *c* axis lattice parameters. The ratio $c_C/c_L$ between *c* axis lattice parameters of the chain and ladder sublattices is approximately equal to 0.7. Thus, here the crystal structure is one-dimensionally modulated and composed of two parts, the chain and ladder sublattices. The structure can also be described as a commensurate structure assuming $10c_C = 7c_L$ for the undoped parent compound using a conventional three dimensional space group approach. However, such a simplified model is not sufficient to describe accurately the crystal structure with doping because the interaction between the two sublattices is strong and doping on Sr site substantially influences the ratio $c_C/c_L$. Moreover, most of the atoms are unrelated by symmetry, resulting in a very high number of free parameters, which are in fact related when using a modulation model. Therefore, the crystal structure should be described as incommensurate using (3+1) dimensional superspace group symmetry, which was introduced by Janner & Janssen in 1980.[10, 16] With this four dimensional formalism, all the observed Bragg reflections of $Sr_{14-x}Ca_xCu_{24}O_{41}$ have to be indexed with four integer indices (*hklm*), where the set of (*hkl0*) corresponds to the reflections from the $CuO_2$ sublattice, the (*hk0m*) are from the ladder sublattice. The (*hklm*) with *lm*≠0 are the intrinsic satellite reflections, corresponding to the mutual structure modulations between the two sublattices.

As mentioned previously, various research groups solved the incommensurate structure of doped/undoped $Sr_{14}Cu_{24}O_{41}$ series compounds by using the superspace model described above. However, different superspace groups were selected, which are reported in TABLE I. If the modulation vector contains more than one component and their denominators have a common divisor, the superspace group contains a so-called non-traditional centering which transforms the unit cell to a basic unit cell. It is denoted using "X" in the superspace group symbols.[17] The space groups discussed in this paper contain the non-traditional centering, thus the non-traditional centering symbols were listed as well in the same table for all the superspace groups. These superspace groups are closely related to each other. Xmmm(0,0,γ)ss0 is the space group with highest symmetry. It changes into X222(0,0,γ)000 by losing the inversion center and changes into Xmma(0,0,γ)ss0 by changing centering operation $T_1$ [(0, ½, ½, ½), (½, 0, ½, ½), (½, ½, 0, 0)] to $T_2$ [(0, ½, ½, ½)]. When Xmma(0,0,γ)ss0 loses the inversion center, it changes to $X2_1ma$(0,0,γ)0s0. For undoped samples, Xmma(0,0,γ)ss0 seems acceptable while published space groups for doped $Sr_{14-x}Ca_xCu_{24}O_{41}$ are contradictive to some experimental results.

Jensen *et al.*[8] chose acentric space group F222 for the Ca doped sample by checking the extinction rules from the single crystal diffraction data. However, this is in contradiction to the initial observation of McCarron et al,[11] who indicated that the small ion Ca doping changes the crystal structure into a higher symmetry confirmed by the transmission electronic microscope (TEM) observation by convergent beam electron diffraction. Both authors concluded that the two sublattices of intermediately Ca doped samples have Fmmm space group. Ohta et al.[3] did not show any reason to select F222 as the space group and did not deny the possibility of centrosymmetric space group. On the other hand, all mentioned authors agreed that the ladder sublattice has an Fmmm space group, hinting the same space group for the chain sublattice. The violation of extinction rules could be attributed to the local twins due to the broken chains.

III. EXPERIMENT

Single crystal samples of spin ladder series $Sr_{14-x}Ca_xCu_{24}O_{41}$ with $x$ =3, 7, 11, 12.2 were grown using the traveling solvent floating zone method under various oxygen pressure up to 15bar.[18] All the samples were made from high-purity starting materials $CaCO_3$, $SrCO_3$ and $CuO$ (purity>99.99%) in order to avoid apparent impurity effects. Small pieces of single crystals with $x$=3 and 12.2 were cut for single crystal X-ray diffraction experiments (Four circle X-ray diffractometer Xcalibur2 with CCD detector, Oxford Diffraction) in order to check their symmetry and space group. The pieces from the good quality part of single crystals were ground into fine powders for neutron powder diffraction experiments. Neutron powder diffraction experiments were carried out on high-resolution diffractometer for thermal neutrons (HRPT, high intensity mode $\Delta d/d \geq 1.8 \times 10^{-3}$) at the SINQ spallation source of Paul Scherrer Institute, Switzerland. The diffraction patterns were collected at 300K and 4K with the wavelength $\lambda$=1.494 and 2.45Å, respectively. The refinements of the crystal structure parameters were performed with the JANA2006 software,[19] which is also used to calculate bond angle, length, bond valence. The modulated structure of $Sr_{14-x}Ca_xCu_{24}O_{41}$ was generated by JANA2006 in an approximate way, and then plotted by VESTA program.[20]

IV. RESULTS AND DISSCUSSIONS
   A. Superspace group and structural refinement

According to single crystal X-ray diffraction in this work, in which only two strong main reflections violated extinction rules, the refinement with Fmmm space group for both subsystems resulted in $R_{wp}$=0.0557 for $Sr_{1.8}Ca_{12.2}Cu_{24}O_{41}$, and 0.0363 for $Sr_{11}Ca_3Cu_{24}O_{41}$, which is quite good and much smaller than the reported values.[7] Thus, we finally adopted Xmmm(00γ)ss0 (No. 69.3 in Tables 9.8.3.5)[14] superspace group for both subsystems to refine the obtained neutron powder diffraction data.

The symmetry operations which have been used are

(1)  $x_1$        $x_2$        $x_3$        $x_4$
(2)  ½-$x_1$      $x_2$        $x_3$        ½+$x_4$
(3)  $x_1$        -$x_2$       $x_3$        ½+$x_4$
(4)  ½+$x_1$      $x_2$        -$x_3$       -$x_4$
(5)  -$x_1$       -$x_2$       -$x_3$       -$x_4$
(6)  ½+$x_1$      -$x_2$       -$x_3$       ½-$x_4$
(7)  -$x_1$       $x_2$        -$x_3$       ½-$x_4$
(8)  ½-$x_1$      -$x_2$       $x_3$        $x_4$

and X-centering translations are (0, 0, 0, 0), (0, ½, ½, ½), (½, ½, 0, 0) and (½, 0, ½, ½,).

The atomic position of atom number $v$ in sublattice $i$ in the modulated composite crystal can be described as[7]

$$\vec{R}_{vi} = \vec{R}^0_{vi} + \sum_{k=1}^{2} \left[ \vec{U}_{sv,k} \cdot \sin(2\pi k \vec{q}_i \cdot \vec{R}^0_{vi}) + \vec{U}_{cv,k} \cdot \cos(2\pi k \vec{q}_i \cdot \vec{R}^0_{vi}) \right] \quad (1)$$

where $R^0_{vi}$ is the position of the atom in the average structure, $k$ is the order of the harmonic, $U_{sv}$ and $U_{cv}$ are three dimensional vectors describing the modulation amplitudes, and $q_i$ is the modulation vector for sublattice $i$. $U_{sv}$ and $U_{cv}$ have to obey symmetry restrictions for atoms on special positions, according to the given superspace symmetry.

At the first steps of the data refinement, the unmodulated incommensurate average structure (including two sublattices: $i$=1, chain sublattice; $i$=2; ladder sublattice) was refined with isotropic temperature factors for all atoms. At this stage e.g. for $Sr_7Ca_7Cu_{24}O_{41}$ sample, R and $R_{wp}$ values were 0.061 and 0.077, respectively. The modulation of the occupation of Sr/Ca and the modulation of the atomic displacement of atom positions have been tried. The former did not contribute to the reduction of the R and $R_{wp}$ values while the latter decreased these two values slightly. With the introduction of the first harmonic modulation of all atomic positions, the refinements converged to an R value of 0.049 and $R_{wp}$ value of 0.062. With the second harmonic modulation of atomic positions, R and $R_{wp}$ shrunk further to 0.045 and 0.059, respectively. In the last step, we refine the data with anisotropic atomic displacement parameters, resulting in substantial decrease of R and $R_{wp}$ (0.0342 and 0.0431, respectively). According to the refinement, $U_{22}$ values were much larger than $U_{11}$ and $U_{33}$ for those atoms on the ladders. And $U_{22}$ and $U_{33}$ values is larger than $U_{11}$ for the atoms on the chains. These results indicate that the atoms do vibrate anisotropically and are consistent with the layer structure of this compound. The atoms on the chains have larger vibration amplitudes than the ones on the ladders. With the isotropic refinement, the temperature factors are all positive. With the anisotropic refinement, a few temperature factors $U_{ii}$ become very small negative values, which were fixed to zero in the final refinement step, however, not strongly influencing the refined results. The final refined parameters and results are listed in Table II for all four samples. The structural model was graphed in Fig. 1(a) and (b) on the basis of the refined results of the $Sr_3Ca_{11}Cu_{24}O_{41}$ sample. As an example, FIG. 2 shows the experimental, calculated and difference neutron powder diffraction profiles of $Sr_7Ca_7Cu_{24}O_{41}$, where Bragg peaks from main reflections, each sublattice and satellites are shown below.

It is noteworthy that all the refinements are based on the same F-centering space group, even for the lowly Ca doped compound $Sr_{11}Ca_3Cu_{24}O_{41}$, which indicates that slightly doping with Ca (*x*=3) is sufficient to change the undoped parent compound from the A centering to F-centering, maintaining to highly Ca doped compounds.

### B. Ca doping effects on the structure

#### i. Lattice parameters

Ca doping strongly influences lattice parameters of both sublattices. Table II presents the refined lattice parameters with the standard deviations for each composition. FIG. 3(a) shows *a* and *b* parameters for Ca doped samples at 300K and 4K. Both decrease linearly with an increase of Ca content. This tendency is reasonable since the ionic radius of Ca is much smaller than that of Sr, and consistent with a previous report on Ca and Ba doped $Sr_{14-x}(Ca, Ba)_xCu_{24}O_{41}$ series by Kato *et al.*[21] It is worthwhile to notice that Ca doping more strongly effects *b* parameter than both *a* and *c*. The former shrinks about 6% going from *x*=3 to 12.2. As shown in FIG. 3(b), $c_C$ shortens about 0.4%, which is much smaller comparing with ~1% in case of $c_L$. The ladder lattice parameter $c_L$ shows an inverse proportional relationship with the Ca content while the chain one $c_C$ exhibits more complicated relationship. Surprisingly at 300K, it increases with the Ca doping for $x \leq 11$, and again decreases at $x=12.2$. At 4K it stays nearly constant at $x \leq 11$, but similarly decreases at higher Ca content. The sublattice parameters are not only dependent on the chemical pressure created by Ca doping, but also on the interaction between the two sublattices. The former has an effect to reduce the lattice parameters while the latter depends how the two sublattices interact each other. It seems that the ladders play a stretching role on the chains in the sample with $x \leq 11$.

The volumes and modulation vectors of the two sublattices are plotted in FIG. 4 (a) and (b), respectively. The volumes are substantially reduced for both sublattices by increasing Ca doping while the modulation vector $q$ increases with doping. $q$ is equal to 0.699 for $x=3$ and ~0.705 for $x=12.2$. The modulation vector of the parent compound $Sr_{14}Cu_{24}O_{41}$ was reported to be 0.699 by Isobe et al.,[13] which is consistent with our low Ca doped sample. However, it is worthwhile to notice that Cox et al.[22] and Etrilard et al.[23] reported much lower q vector for $Sr_{14}Cu_{24}O_{41}$. This can be attributed to the different sample preparation since the oxygen content in this compound was proved to be critically influential on the structure, especially on the $q$ vector.[24] Additionally, the curve looks saturated at high doping, which may hint that the crystal structure is approaching a certain limit of substitution Sr with Ca. Further Ca doping may cause different type of structural distortion, which is more difficult to implement. For example, crystals with higher Ca content need to be grown in much higher oxygen pressure.

ii. Static and dynamic atomic displacement

The typical feature of modulated structure is that atoms in both sublattices are not fixed at average positions but that are statically displaced according to the equation (1). Moreover, atoms are also displaced dynamically as described by atomic displacement parameters (ADP), formally named as temperature parameters. In this system, the ADP and modulation displacements of each atom are strongly anisotropic because of the layered structure. Our refinement results show that the atoms, especially atoms on the chains and Sr/Ca, have much larger ADP along $b$ axis than along $a$ and $c$ axes. The modulated displacements along $b$ axis show similar features. Here, we only discuss the atomic displacements along *b* direction. FIG. 5 shows the displacements of Sr(1), O(1), Cu(2) and O(2) along *b* axis in the *t* space. The displacements of all these atoms are at the order of magnitude of 0.1Å. However, the modulation of the O(1) have several times larger displacement than the others. Ca doping slightly influences the displacements of each atoms. For example, Ca doping reduces the modulation amplitude of Sr/Ca but increases the modulations of the Cu and O. Since the volumes of both subsystems decrease on doping, the free space in the lattice is reduced and the movements of atoms are limited. However, oxygen displacements increase with Ca doping because of the increase of the interaction between two sublattices. For example, Cu on ladders can bond with oxygen on the chains when increasing *x*, which may contribute to oxygen displacements.

iii. Cu-O bond length

For the convenience of discussion, the chain and ladder structures are plotted with atom labels in FIG. 6. FIG. 7 (a) shows the Cu-O bond length on the chains, which has a positive correlation with Ca doping at both 300K and 4K. This is an important effect of Ca doping since the Cu-O distance directly determines Cu-O bonds and Cu valences. The $CuO_4$ planar units on *the chains* underwent several changes on Ca doping. The $CuO_2$ square, which actually is an elongated rectangle shape along *c* axis, became closer to a regular square shape by shortening along *c* axis when doping with Ca. On the other side, oxygen on the chains modulates larger and larger along *b* axis. Thus, these two effects result in the increase of the Cu-O bond length. FIG. 8 (a) plots the Cu-O distances on the chains modulated in the *t* space. The shortest Cu-O distance does not change much while the modulation amplitude increases with the increase of Ca content, hinting that the interaction between the two sublattices becomes stronger. Such a strong modulation also means the strong localization of the doped holes.

FIG. 7 (b) shows the average Cu-O bond lengths on the ladders at 300K and 4K. As shown in FIG. 6, there are two O sites: O(2) and O(3), on the ladder. Consequently, there are three different Cu-O bond lengths: Cu-O bond length on rungs (i.e. Cu-$O_R$), Cu-O bond length along legs (i.e. Cu-$O_{L1}$); Cu-O bond length between legs (i.e. Cu-$O_{L2}$). In FIG. 7 (b), Cu-$O_{L1}$ is longer than Cu-$O_{L2}$ and Cu-$O_R$ in all the samples. With introduction of more Ca in the system, Cu-$O_{L1}$ and Cu-$O_R$ become shorter and Cu-$O_{L2}$ elongates. Namely, the distance between two ladders become larger and the rung and leg lengths shrink by increasing Ca content. These changes will influence exchange interactions between neighboring $Cu^{3+}/Cu^{2+}$ and the hopping of holes along different axes. We will discuss this in detail in the next section. FIG. 9(a) and (b) plot the modulated Cu-O bond lengths on legs and rungs in the $t$ space, respectively. Similar $t$ dependence of Cu-O distance has been shown by Isobe et al.[25] for $Sr_{0.4}Ca_{13.6}Cu_{24}O_{41}$. The modulation amplitudes of Cu-$O_{L1}$ distance increase with $x$ while the ones of Cu-$O_{L2}$ are not linearly dependent on the Ca doping. The Cu-O distance along rungs (i.e. half rung length) in FIG. 10 shows much weak modulation and decrease clearly with the increase of Ca content.

Another important feature is that the distance between the two sublattices is substantially reduced with Ca substitution for Sr. In FIG. 8(b), the Cu(2)-O(1) distance modulates in the $t$ space and shows several minimal values in one period. The modulated curves are shifted down by increasing Ca content. The minimal Cu(2)-O(1) distance is around 2.81Å, observed in the samples with $x>11$. This value is much shorter than the reported value 3.05 Å for the undoped parent compound. Such a short distance is believed to contribute to Cu bond valences, which will be discussed in the next part.

### iv. Cu-O bond valence

The Cu-O distance directly determines the Cu-O bond valence, which is closely related to charge transfer and spin dynamic behavior in this system. Cu sites on two sublattices are nonequivalent in this material and their bond valence can be calculated by bond valence sum method. The Cu-O reference distance (1.679Å) for $Cu^{2+}$ was chosen from the value published by Brown et al.[26]

FIG. 11(a) shows the average bond valence sums of Cu on the chains, on the ladders and the bond valence of Cu(2)-O(1) at 4K and 300K. For comparison, the bond valence values of the undoped compound published by Gotoh et al.[13] and the velues of highly Ca doped compound with $x$=13.6 by Isobe et al.[25] are plotted as well in the figure as references. It is clear that Cu bond valences on the ladders and chains show opposite tendencies when increasing Ca content. The Cu bond valence sum on the chains decreases gradually from 2.55 to 2.18 between $x$=0 and 12.2. If only taking oxygen on the ladder into account, the Cu bond valence sum on the ladder sites increases from 2.01 to 2.12 within $x = 0$ and 12.2. Additionally, the contribution from O(1) on chain sites increases with $x$. If one takes O(1)'s contribution into account as well, the Cu bond valence sum reaches 2.16 on the ladder at $x$=12.2. This value is very close to the one (2.20) estimated by optical conductivity measurements.[6] More comprehensive discussion about the charge transfer and hole number per formula unit will be presented in details in the coming section.

FIG. 12 shows the Cu bond valence sum modulated with $t$. The curves for the chains oscillate strongly and the amplitude increases with the increase of Ca content. This means that the holes sitting on the chains become even more localized upon doping. For the ladders, on the contrary, the modulation amplitude became smaller and smaller with Ca doping, indicating the doped holes become more and more itinerant. Therefore, the decrease of resistivity on doping is not only due to charge transfer from the chains to ladders, but also due to the delocalization of doped holes on the ladders. FIG. 13 shows the bond valence of Cu(2)-O(1) in the $t$ space. The curves move up when increasing $x$, exhibiting the same tendency as the lowest line in FIG. 11 (a).

v. Cu-O bond angles

The interaction of the two sublattices has strong influence on the bond angles. FIG. 14 shows the average Cu-O angles on the chains and ladders at 300K and 4K. The two angles on chains, $O_A$-Cu-$O_B$ and $O_B$-Cu-$O_C$, show opposite dependences on Ca content. The former increases while the latter decreases with $x$. Both of them approach 90°, which means that $CuO_4$ units become closer to the square shape. In the ladders O-Cu-O angles exhibit a similar but more evident feature (see FIG. 14(b)). With increase of $x$ $O_{L1}$-Cu-$O_{L2}$ and $O_R$-Cu-$O_{L1}$ are approaching ~90°. It is interesting to note that the observed changes are more evident at 4K, indicating that the system becomes closer to orthogonal at low temperature. The modulation curves of these angles in the $t$ space show how these angles evolve with doping in FIG. 15. The modulation of angle $O_R$-Cu-$O_{L2}$ in the $t$ space is shown in FIG. 16. Angle $O_R$-Cu-$O_{L2}$ deviates strongly from 180° near $t = 0.5*n$ (n is an integer) in the samples with $x < 12.2$, but much weaker in the sample with $x =12.2$. This means that O-Cu-O bonds become nearly straight along $a$ axis, which is favorable for carriers to hop from one ladder to the other.

C. Discussion

From the results shown above, Ca substitution for Sr brought drastic changes to the modulated structure of the $Sr_{14-x}Ca_xCu_{24}O_{41}$ system. The unit cell shrunk along *a, b, c* axis. Especially, around 6% shrinkage along *b* axis caused strong interaction between the chains and ladders, which is believed to play indispensible roles in the superconducting transition of the system under external pressure. The second apparent change is that the modulation vector *q* increases from 0.699 to 0.705 upon Ca doping. From our data, *q* seems to saturate for high Ca content. The value of *q* in the $Sr_{0.4}Ca_{13.6}Cu_{24}O_{41}$ powder sample was refined to be 0.70122 by T. Ohta et al.,[3] which is smaller than our observed values (~0.705) in both *x*=11 and 12.2 samples. This seems to indicate that the modulation vector reaches the maximum value at *x*~12. Another related observation is that the *c* axis of the chain sublattice starts to shrink at nearly the same *x* value. The existence of such a maximum may hint that Ca doping has two different effects which are differently pronounced depending on the doping level. It is worth noting that the superconducting state can be induced only in the compounds with *x*>10 additionally applying external pressure. Since this overlaps with the *x* range where the modulation vector saturates, it strongly supports an idea that these two phenomena are directly related. At low Ca doping (*x*<11), most of the structural parameters shows a linear relationship with *x*. When *x* is higher than 11 the angles O(2)-Cu(2)-O(2) and O(2)-Cu(2)-O(3), show a deviation from the linear relationship. Thus, it is reasonable to discuss the role of Ca doping in these two ranges. Isobe et al.[9] pointed out that Ca substitution and external physical pressure differently contributed to the superconductivity in this system. Undoubtedly, Ca doping, especially at the level below *x*=11, increases the carrier concentration on the ladder, which could not be realized by applying physical pressure only. This claim is supported by the fact that $Sr_{14}Cu_{24}O_{41}$ is not a superconductor under pressure up to 8GPa.[13] However, at *x*>11, Ca doping not only introduces the shrinkage of the lattice, but also substantially adjusts the Cu-O bond angles and distances. The decrease of modulation vector hints interlayer interaction in a different way. For 11<*x*<13.6, the optimal physical pressure for highest $T_C$ decreases with increase of *x*, which may hint that further Ca doping in this range behaves similarly as the physical external pressure.

On Ca doping, *q* increases from 0.699 to 0.705, resulting in the increase in the $CuO_2/Cu_2O_3$ ratio. This means that the hole concentration increases of ~0.07 per formula. Ca doping increases the modulation of the chain, causing more localized holes on chains. Therefore, the ladder, which has more non-localized holes on doping, contributes to the main part of conductivity. Isobe *et al.*[9] assumed that Ca doping strongly increase coupling between the neighboring ladders. However, oppositely our results show that the Cu-O distances between two neighboring ladders increase upon Ca doping. This means that the interladder exchange coupling becomes weaker. It should be noted that the Cu-O distance on rungs also decrease upon Ca doping, which could favor the conduction along *a* axis. In addition, O-Cu-O bond angles become orthogonal or straight with increasing Ca doping. It is known that deviation of O-Cu-O angle from 180$^{o}$ reduces the superconducting transition temperature for $La_{2-x}Sr_xCuO_4$.[27] Similarly in highly Ca doped $Sr_{14-x}Cu_xCu_{24}O_{41}$, the decrease of deviation of the O-Cu-O bond straightness may be one important feature for the superconducting transition.

Phase diagrams of cuprate superconductors show that an appropriate carrier doping is necessary for a system to become superconducting. The most widely used way is to dope the system with cations with different valences. Charge transfer is another way to dope hole or electron into superconducting $CuO_2$ planes. In this case, there is an electron reservoir next to the superconducting plane providing doping charges to the superconducting plane.

The charge transfer was believed to be an essential modification for the appearance of the superconductivity in the highly doped $Sr_{14-x}Ca_xCu_{24}O_{41}$ samples. There are two open questions about the charge transfer in this spin ladder series. The first is how many charges are transferred from the chains to the ladders with the Ca doping. FIG. 11(b) summarizes the research results about the hole number of different Ca doped compounds, which were measured by different groups using different methods, including optical spectroscopy[6], nuclear magnetic resonace,[28] X-ray absorption spectroscopy[29] and so on. However, the discrepancies among them are quite obvious from one to the other. The most systematic measurements have been done by Osafune *et al.*[6] using optical conductivity method, by Nücker *et al.*[30] using polarization-dependent near-edge x-ray absorption fine structure (NEXAFS) method, and by us with the BVS method. Nücker's results indicate much less charge transfer from chains to ladders with Ca doping than the optical and BVS results. The Osafune's optical method is based on the assumption of 6 holes per formula and respectively calculates the chain and ladder contributions by their optical responses, namely, the so-called spectral weight. Since the optical results were based on the assumption of 6 holes per formula, we normalized our BVS data with the same assumption, as the red curves shown in FIG. 11(b). Carefully comparing the normalized data and the results from Osafune *et al.*,[6] we could find that these two sets of data are very similar, especially for highly Ca doped compositions. It should not be ignored the data from Rusydi *et al.*,[29] which show much higher hole number on the ladders, among which the parent compound has very similar valences for Cu on chains and ladders. A comprehensive analysis should be carried out in order to sort out this controversy. It is more reasonable to take the whole series $(La,Y)_y Sr_{14-y} Cu_{24} O_{41}$ into account, where the doped holes are controlled by the doped trivalent ions La or Y. The magnetic susceptibility measurements revealed that the magnetic behavior gradually changed with the doped hole number until the undoped sample $La_6 Sr_8 Cu_{24} O_{41}$.[31] Electrical transport results by T. Ivek *et al.*[32] disclosed that holes mainly sits on the chain sites and distributed to the ladder sites only when hole number is larger than 4 in the whole system. They estimated that the sample with y=1 and 2 have 5 and 4 holes on chains, respectively. Thus, we should safely expect higher holes number in the parent compound. This agrees with all the reported data in FIG. 11(b) except Rusydi's result. Additionally, the manifest differences between chains and ladders on this compound have been observed by optical conductivity,[6] NMR,[28] NEXAFS[30] and BVS calculations in this work. This fact could not be understood in the frame of Rusydi's model. On the basis of the consistence of all these data discussed above, the conclusion of high proportional holes on ladders made by Rusydi *et al.*[29] seems quite unconvincing, at least needs some further qualification.

The second open question is about the mechanism of charge transfer. According to Ohta *et al.*[3] and Gotoh *et al.*,[13] charge transfer is mainly caused by the apical oxygen, which is located on the chains and moves to the ladder due to the reduction of the interlayer distance by small radius element Ca substitution for Sr. This explanation was derived from the mechanism observed in YBCO. However, as presented by these authors, the bond valence contribution from the apical oxygen to the system is quite limited, only around 0.03. Furthermore, it has to be noticed that the distance from Cu to the apical oxygen is 2.8Å in the present compounds, which is much larger than this value (~2.3Å) for YBCO.[33] Even under 7GPa pressure,[9] the $Cu_{ladder}$-$O_{apical}$ distance in spin ladder compound could not catch up with YBCO. However, the linear R-$T^2$ relationship observed in $Sr_{14-x} Ca_x Cu_{24} O_{41}$ indicated that this material is overdoped.[34] The doped holes is estimated to be ~0.2 per Cu on the ladders in the sample *x*=12 under ambient pressure from optical conductivity measurement conducted by Osafune *et al.*[6] Thus, the apical oxygen cannot be assumed to provide such a high concentration of carriers. There must be some other sources for the charge transfer. Different from YBCO, $Sr_{14-x}Ca_xCu_{24}O_{41}$ is a modulated structure, in which the sublattices are distorted due to the strong interaction between chains and ladders. According to our calculation, in addition to the contribution from the apical oxygen, the modulated structural changes caused by Ca substitution played a crucial role in the charge transfer from the hole reservoir in the chains to the ladder. The average doped holes per Cu on the ladders are around 0.1. Taking both effects into account, there is around 0.16 increase in Cu valence for the sample with $x$=12.2. This value is quite close to the value estimated by optical measurement. Therefore, our experimental data confirms the observation of the optical conductivity measurement and provides structural evidence for the charge transfer from the chains to the ladders.

The coordination shell of Cu on ladder is shown in FIG. 1(b). The apical oxygen atoms create pairs along the rungs, but they don't approach Cu from the same but opposite sides of the ladders. This is unlike the model proposed by Isobe *et al.*,[35] in which the apical oxygen paired side by side along ladder and connected by rung from the same side of the ladder. Thus, our model has much less local distortion than Isobe's model.

Comparing the results at 300K and 4K, there is no large difference between them. It is noteworthy that Cu valence sum on the chains increased at low temperature while it did not change much on the ladders. This result is consistent with Isobe's observation that charges partially flow back to the chains at low temperature.[25] In addition, O-Cu-O bond angles on the ladder approach 90º with Ca doping at 4K much quicker than at 300K. This suggests that the suppression of lattice distortion is essential for the superconductor transition.

The spin gap theoretically predicted for even-leg spin ladders were observed for this system. The spin gap become wider upon increasing $J'/J$, where $J'$ and $J$ stand for Cu-Cu exchange coupling energy along the rungs and legs of the ladder, respectively.[36] According to our observation, both Cu-Cu distances along the rungs and legs shrunk upon Ca doping, and the rung distance decrease slightly quicker than that along leg, especially at low temperature, where the magnetic excitation becomes obvious. This may be the reason why we observed the spin gap in the ladder slightly shifts to higher energy on doping by inelastic neutron scattering. The results (not shown here) showed the spin gap at 32meV and 33.5meV in the samples with $x$=0 and 3, respectively.[37]

## V. CONCLUSION

Neutron powder diffraction experiments were conducted on the spin ladder series $Sr_{14-x}Ca_xCu_{24}O_{41}$ ($x$=3, 7, 11, 12.2). An incommensurate modulated structural model with a superspace group Xmmm(00γ)ss0 was used to fit these data, resulting in a good agreement between the theoretical and experimental data. The refined crystal structure results reveal the evolution of the modulation between the chain and ladder sublattice with the increase of Ca substitution for Sr. On Ca doping, the lattice parameters shrink as expected, and the modulation vector increases from 0.699 to 0.705 for the samples with $x$ from 3 to 12.2. Ca doping not only has a strong effect on the Cu-O bond lengths and O-Cu-O angles on the chains and ladders, but also increases the interaction between the two chain and ladder sublattices. The bond valence sum calculation provides clear evidence for charge transfer from the chains to the ladders with increasing doping, which is primarily responsible for the electrical transport behavior observed in $Sr_{14-x}Ca_xCu_{24}O_{41}$.

**Acknowledgement:** The authors gratefully acknowledge the financial support from the Indo-Swiss Joint Research Program (ISJRP, Contract No. JRP122960) by Swiss State Secretariat of Education and Research. The authors would like to thank also Dr. Michal Dusek and Dr. Karla Fejfarova from Institute of Solid State Physics for checking samples by X-ray diffraction experiments. V. Petříček thanks support from Praemium Academiae of the Czech Academy of Sciences.

**References**

1    E. Dagotto, J. Riera, and D. Scalapino, Phys Rev B **45**, 5744 (1992).
2    T. Vuletic, B. Korin-Hamzic, T. Ivek, S. Tomic, B. Gorshunov, M. Dressel, and J. Akimitsu, Phys Rep **428**, 169 (2006).

3   T. Ohta, F. Izumi, M. Onoda, M. Isobe, E. Takayama-Muromachi, and A. W. Hewat, J Phys Soc Jpn **66**, 3107 (1997).
4   L. P. Regnault, J. P. Boucher, H. Moudden, J. E. Lorenzo, A. Hiess, U. Ammerahl, G. Dhalenne, and A. Revcolevschi, Phys Rev B **59**, 1055 (1999).
5   K. I. Kumagai, S. Tsuji, M. Kato, and Y. Koike, Phys Rev Lett **78**, 1992 (1997).
6   T. Osafune, N. Motoyama, H. Eisaki, and S. Uchida, Phys Rev Lett **78**, 1980 (1997).
7   A. F. Jensen, F. K. Larsen, B. B. Iversen, V. Petricek, T. Schultz, and Y. Gao, Acta Crystallogr B **53**, 113 (1997).
8   A. F. Jensen, V. Petricek, F. K. Larsen, and E. M. McCarron, Acta Crystallogr B **53**, 125 (1997).
9   M. Isobe, T. Ohta, M. Onoda, F. Izumi, S. Nakano, J. Q. Li, Y. Matsui, E. Takayama-Muromachi, T. Matsumoto, and H. Hayakawa, Phys Rev B **57**, 613 (1998).
10  P. M. d. Wollf, T. Janssen, and A. Janner, Acta Crystallogr A **37**, 625 (1981).
11  E. M. Mccarron, M. A. Subramanian, J. C. Calabrese, and R. L. Harlow, Mater Res Bull **23**, 1355 (1988).
12  K. Ukei, T. Shishido, and T. Fukuda, Acta Crystallogr B **50**, 42 (1994).
13  Y. Gotoh, I. Yamaguchi, Y. Takahashi, J. Akimoto, M. Goto, M. Onoda, H. Fujino, T. Nagata, and J. Akimitsu, Phys Rev B **68**, 224108 (2003).
14  T. Janssen, A. Janner, A. Looijenga-Vos, and P. M. d. Wolff, International Tables for Crystallogr Vol C, 907 (2006).
15  V. K. Kato, Acta Crystallogr B **46**, 39 (1990).
16  A. Janner and T. Janssen, Acta Crystallogr A **36**, 408 (1980).
17  I. Orlov, L. Palatinus, and G. Chapuis, J Appl Cryst **41**, 1182 (2008).
18  G. Deng, D. M. Radheep, R. Thiyagarajan, E. Pomjakushina, S. Wang, N. Nikseresht, S. Arumugam, and K. Conder, J Crystal Growth **327**, 182 (2011).
19  V. Petricek, M. Dusek, and L. Palatinus, Jana2006 - the crystallographic computing system. Institute of Physics of the ASCR, Prague, Czech Republic, www-xray.fzu.cz/jana (2006).
20  K. Momma and F. Izumi, J Appl Cryst **41**, 653 (2008).
21  M. Kato, K. Shiota, and Y. Koike, Physica C **258**, 284 (1996).
22  D. E. Cox, T. Iglesias, K. Hirota, G. Shirane, M. Matsuda, N. Motoyama, H. Eisaki, and S. Uchida, Phys Rev B **57**, 10750 (1998).
23  J. Etrillard, M. Braden, A. Gukasov, U. Ammerahl, and A. Revcolevschi, Physica C **403** 290 (2004).

[24] Z. Hiroi, S. Amelinckx, G. V. Tendeloo, and N. Kobayashi, Phys Rev B **54**, 15849 (1996).

[25] M. Isobe, M. Onoda, T. Ohta, F. Izumi, K. Kimoto, E. Takayama-Muromachi, A. W. Hewat, and K. Ohoyama, Phys Rev B **62**, 11667 (2000).

[26] I. D. Brown and D. Altermatt, Acta Crystallogr B **41**, 244 (1985).

[27] C. Rial, E. Morfin, M. A. Alario-Franco, U. Amador, and N. H. Andersen, Physica C **254**, 233 (1995).

[28] Y. Piskunov, D. Jérome, P. Auban-Senzier, P. Wzietek, and A. Yakubovsky, Phys Rev B **72**, 064512 (2005).

[29] A. Rusydi, M. Berciu, P. Abbamonte, S. Smadici, H. Eisaki, Y. Fujimaki, S. Uchida, M. Rübhausen, and G. A. Sawatzky, Phys Rev B **75** (2007).

[30] N. Nucker, M. Merz, C. A. Kuntscher, S. Gerhold, S. Schuppler, R. Neudert, M. S. Golden, J. Fink, D. Schild, S. Stadler, V. Chakarian, J. Freeland, Y. U. Idzerda, K. Conder, M. Uehara, T. Nagata, J. Goto, J. Akimitsu, N. Motoyama, H. Eisaki, S. Uchida, U. Ammerahl, and A. Revcolevschi, Phys Rev B **68**, 224108 (2003).

[31] N. Motoyama, T. Osafune, T. Kakeshita, H. Eisaki, and S. Uchida, Phys Rev B **55**, R3386 (1997).

[32] T. Ivek, T. Vuletic, B. Korin-Hamzic, O. Milat, and S. Tomic, Phys Rev B **78**, 205105 (2008).

[33] P. Schweiss, W. Reichardt, M. Braden, G. Collin, G. Heger, H. Claus, and A. Erb, Phys Rev B **49**, 1387 (1994).

[34] T. Nagata, M. Uehara, J. Goto, J. Akimitsu, N. Motoyama, H. Eisaki, S. Uchida, H. Takahashi, T. Nakanishi, and N. Mori, Phys Rev Lett **81**, 1090 (1998).

[35] M. Isobe, T. Ohta, M. Onoda, F. Izumi, S. Nakano, J. Q. Li, Y. Matsui, E. Takayama-Muromachi, T. Matsumoto, and H. Hayakawa, Phys Rev B **57**, 613 (1998).

[36] T. Barnes, E. Dagotto, J. Riera, and E. S. Swanson, Phys Rev B **47**, 3196 (1993).

[37] G. Deng, Unpublished results from LLB (2009).

TABLE I. Comparison of different superspace groups used for the structural refinement by various authors. The symbols of traditional superspace group, superspace group for the unified unit cell and the non-traditional centering are listed as well. The relationships between these superspace groups are explained in the text in detail.

| Source | K. Ukei et al.[12] | Y. Gotoh et al.[13] | A. F. Jensen et al.[8] | A. F. Jensen et al.[7] | M. Isobe et al.[9] | This work |
|---|---|---|---|---|---|---|
| Sample | $Sr_{14}Cu_{24}O_{41}$ | $Sr_{14}Cu_{24}O_{41}$ | $Bi_{0.56}Sr_{13.44}Cu_{24}O_{41}$ | $Sr_{6.44}Bi_{0.84}Ca_{6.72}Cu_{24}O_{41}$ | $Sr_{0.4}Ca_{13.6}Cu_{24}O_{41}$ | $Sr_{14-x}Ca_xCu_{24}O_{41}$ |
| Published symbol | $A2_1ma(0,1,1/\gamma)0s0$ | $Amma(0,1,\gamma)ss0$ | $Amma(0,0,1+\gamma)ss0$ | $F222(0,0,1+\gamma)000$ | $F222(0,0,\gamma)000$ | $Fmmm(0,0,1+\gamma)ss0$ |
| $q_x$ | 0 | 0 | 0 | 0 | 0 | 0 |
| $q_y$ | 1 | 1 | 0 | 0 | 0 | 0 |
| $q_z$ | $1/\gamma$ | $\gamma$ | $1+\gamma$ | $1+\gamma$ | $\gamma$ | $1+\gamma$ |
| Inversion Centering | acentric | centrosymmetric | centrosymmetric | acentric | acentric | centrosymmetric |
| Symbol of unified cell | $A2_1ma(0,0,1+\gamma)0s0$ | $Amma(0,0,1+\gamma)ss0$ | $Amma(0,0,1+\gamma)ss0$ | $F222(0,0,1+\gamma)000$ | $F222(0,0,1+\gamma)000$ | $Fmmm(0,0,1+\gamma)ss0$ |
| X-centering symbol | $X2_1ma(0,0,\gamma)0s0$ | $Xmma(0,0,\gamma)ss0$ | $Xmma(0,0,\gamma)ss0$ | $X222(0,0,\gamma)000$ | $X222(0,0,\gamma)000$ | $Xmmm(0,0,\gamma)ss0$ |

TABLE II. The structure parameters of $Sr_{14-x}Ca_xCu_{24}O_{41}$ (x=3, 7, 11, 12.2) composite compounds at 300K. The data are refined from the powder neutron diffraction patterns measured at HRPT/SINQ with wavelength λ=1.494 Å. $x$, $y$, $z$ are fractional atomic coordinates. $O_{cc}$ is the site occupancy. $U_{eq}$ is the atomic temperature parameter (=$B/8/\pi^2$), which is given in Å$^2$. $U_{si}$, $U_{ci}$ are the modulation amplitude of the sine and cosine wave given in Å, respectively. Cu(1) and O(1) belong to chain sublattice (i=1). Cu(2), O(2), O(3) and Sr/Ca belong to ladder sublattice (i=2).

| Ca content | x=3 | x=7 | x=11 | x=12.2 |
|---|---|---|---|---|
| a | 11.4221(3) | 11.3627(3) | 11.3007(3) | 11.2786(3) |
| b | 13.2021(4) | 12.9151(4) | 12.6204(4) | 12.5244(5) |
| $c_C$ | 2.7450(3) | 2.7477(2) | 2.7480(2) | 2.7459(2) |
| $c_L$ | 3.9261(4) | 3.9103(3) | 3.8984(3) | 3.8948(3) |
| q | 0.69917 | 0.70269 | 0.70490 | 0.704937 |
| $V_C/V_L$ | 413.9/592 | 403.2/573.8 | 391.9/556.0 | 387.8/550 |
| Cu(1) (x,y,z) | 0.5,0.5,0 | 0.5,0.5,0 | 0.5,0.5,0; | 0.5,0.5,0; |
| $O_{cc}$;$U_{eq}$ | 0.5;0.045(2) | 0.5;0.0330(15) | 0.5;0.0210(14) | 0.5;0.0185(11) |
| $U_{sx},U_{sy},U_{sz}$(q) | 0,0,0 | 0,0,0 | 0,0,0 | 0,0,0 |
| $U_{cx},U_{cy},U_{cz}$(q) | 0,0,0 | 0,0,0 | 0,0,0 | 0,0,0 |
| $U_{sx},U_{sy},U_{sz}$ (2q) | 0,0, -0.003(3) | 0,0, 0.003(2) | 0,0,0.001(2) | 0,0,0.0035(19) |
| $U_{cx},U_{cy},U_{cz}$ (2q) | 0,0,0 | 0,0,0 | 0,0,0 | 0,0,0 |
| O(1) (x,y,z) | 0.8885(2),0.5,0 | 0.88663(17),0.5,0 | 0.88512(2),0.5,0 | 0.88422(18),0.5,0; |
| $O_{cc}$;$U_{eq}$ | 0.25;0.0324(17) | 0.25;0.0335(15) | 0.25;0.0289(16) | 0.25;0.0333(17) |
| $U_{sx},U_{sy},U_{sz}$ (q) | 0,0,0 | 0,0,0 | 0,0,0 | 0,0,0 |
| $U_{cx},U_{cy},U_{cz}$ (q) | 0,-0.0207(5),0 | 0,-0.0228(5),0 | 0,-0.0238(5),0 | 0,-0.0242(6),0 |
| $U_{sx},U_{sy},U_{sz}$ (2q) | 0,0, 0.000(3) | 0,0, 0.0127(19) | 0,0,0.016(3) | 0,0,0.027(2) |
| $U_{cx},U_{cy},U_{cz}$ (2q) | -0.0027(5),0,0 | -0.0014(4),0,0 | -0.0035(5),0,0 | -0.0033(5),0,0 |
| Sr/Ca(x,y,z) | 0.25,0.3785(2),0 | 0.25,0.37929(18),0 | 0.25,0.3800(3),0 | 0.25,0.3794(2),0 |
| $O_{cc}$ | 0.392(8)/0.108(8) | 0.248(7)/0.252(7) | 0.101(8)/0.398(8) | 0.075(8)/0.418(8) |
| $U_{eq}$ | 0.0096(11) | 0.0106(9) | 0.0085(9) | 0.0076(9) |
| $U_{sx},U_{sy},U_{sz}$ (q) | 0,0,0 | 0,0,0 | 0,0,0 | 0,0,0 |
| $U_{cx},U_{cy},U_{cz}$ (q) | 0.0041(11),0,0 | 0.0014(10),0,0 | 0.0019(9),0,0 | 0.0026(13),0,0 |
| $U_{sx},U_{sy},U_{sz}$ (2q) | 0,0, 0.007(3) | 0,0,-0.010(3) | 0,0,-0.016(3) | 0,0,-0.008(3) |
| $U_{cx},U_{cy},U_{cz}$ (2q) | 0,0.0072(16),0 | 0, 0.0063(14),0 | 0,-0.0038(16),0 | 0,-0.0042(14),0 |
| Cu(2)(x,y,z) | 0.58422(16),0.25,0 | 0.58439(2),0.25,0 | 0.58415(16),0.25,0 | 0.58421(16),0.25,0; |
| $O_{cc}$;$U_{eq}$ | 0.25;0.0102(6) | 0.25;0.0109 (6) | 0.25;0.0061(6) | 0.25;0.0078(6) |
| $U_{sx},U_{sy},U_{sz}$ (q) | 0,0,0 | 0,0,0 | 0,0,0 | 0,0,0 |
| $U_{cx},U_{cy},U_{cz}$ (q) | 0,-0.0049(10),0 | 0,-0.0051(6),0 | 0,-0.0069(6),0 | 0,-0.0039(14),0 |
| $U_{sx},U_{sy},U_{sz}$ (2q) | 0,0,0.018(2) | 0,0, 0.0118(18) | 0,0,-0.003(2) | 0,0,0.006(2) |
| $U_{cx},U_{cy},U_{cz}$ (2q) | 0.0007(11),0,0 | 0.0003(11),0,0 | 0.0004(10),0,0 | 0.0010(10),0,0 |
| O(2) (x,y,z) | 0.4198(2),0.25,0 | 0.41815(18),0.25,0 | 0.4169(2),0.25,0 | 0.4167(2),0.25,0; |
| $O_{cc}$;$U_{eq}$ | 0.25;0.0126(9) | 0.25;0.0112(6) | 0.25;0.0101(9) | 0.25;0.0098(7) |
| $U_{sx},U_{sy},U_{sz}$ (q) | 0,0,0 | 0,0,0 | 0,0,0 | 0,0,0 |
| $U_{cx},U_{cy},U_{cz}$ (q) | 0,-0.0002(12),0 | 0, -0.0018(9),0 | 0,-0.0021(8),0 | 0,-0.0065(9),0 |
| $U_{sx},U_{sy},U_{sz}$ (2q) | 0,0,-0.022(3) | 0,0,-0.026(3) | 0,0,-0.022(3) | 0,0,-0.015(3) |
| $U_{cx},U_{cy},U_{cz}$ (2q) | 0.0028(17),0,0 | -0.0029(11),0,0 | -0.0038(15),0,0 | -0.0042(12),0,0 |
| O(3) (x,y,z) | 0.25,0.25,0.5 | 0.25,0.25,0.5 | 0.25,0.25,0.5 | 0.25,0.25,0.5; |
| $O_{cc}$;$U_{eq}$ | 0.5;0.0057(13) | 0.5;0.0142(11) | 0.5;0.0133(13) | 0.5;0.0149(12) |
| $U_{sx},U_{sy},U_{sz}$ (q) | 0,0,0 | 0,0,0 | 0,0,0 | 0,0,0 |
| $U_{cx},U_{cy},U_{cz}$ (q) | 0,0,0 | 0,0,0 | 0,0,0 | 0,0,0 |
| $U_{sx},U_{sy},U_{sz}$ (2q) | 0,0, 0.028(4) | 0,0,0.020(3) | 0,0,0.010(3) | 0,0,0.020 (6) |
| $U_{cx},U_{cy},U_{cz}$ (2q) | 0,0,0 | 0,0,0 | 0,0,0 | 0,0,0 |
| $R_{wp}$, $R_{exp}$, GOF | 3.49, 4.57,1.81 | 3.35, 4.20, 1.48 | 3.84, 4.96, 1.66 | 3.80, 4.79, 1.57 |
| Refined Formula | $Sr_{10.99}Ca_{3.01}Cu_{24}O_{41.00}$ | $Sr_{8.28}Ca_{5.74}Cu_{24}O_{40.99}$ | $Sr_{2.85}Ca_{11.19}Cu_{24}O_{40.98}$ | $Sr_{2.12}Ca_{11.74}Cu_{24}O_{40.98}$ |

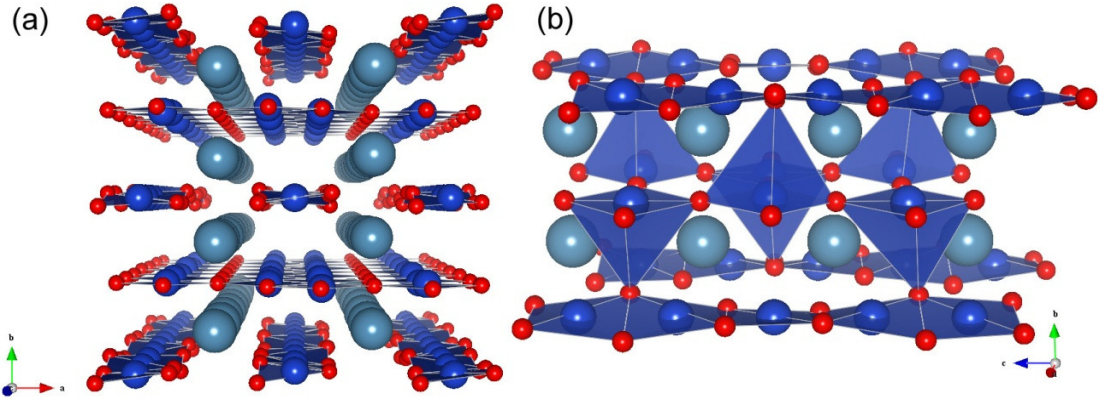

FIG. 1. Typical incommensurate modulated structure of $Sr_{14-x}Ca_xCu_{24}O_{41}$ (a) and partial structure showing $CuO_5$ pyramid coordinates of Cu(2) with the apical O1 on the chains (c). The blue spheres denote Cu atoms, the red small spheres denote O atoms and the biggest spheres denote Sr/Ca atoms.

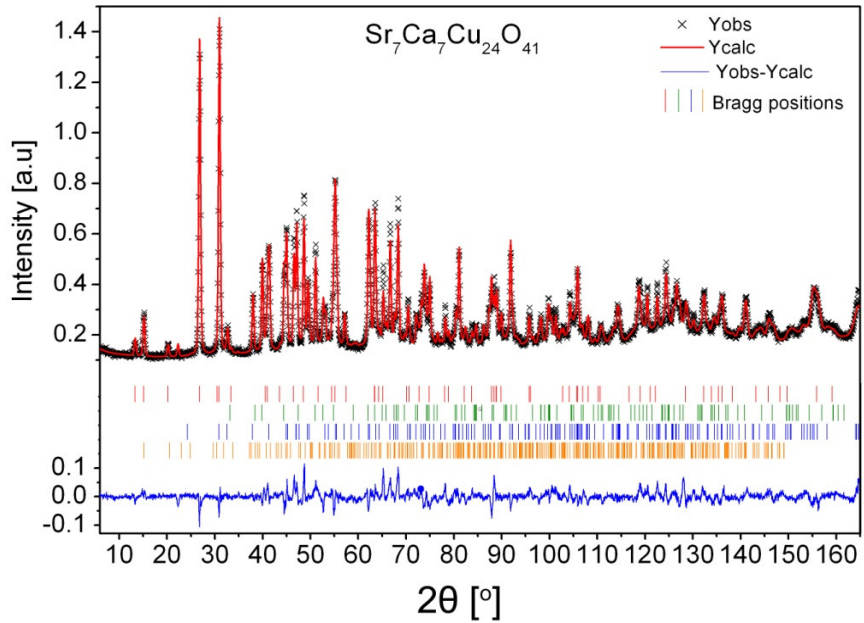

FIG. 2. The experimental (black crosses), calculated (red curve) and difference neutron powder diffraction profiles ($\lambda$=1.494Å, T=300K) for the powderized $Sr_7Ca_7Cu_{24}O_{41}$ single crystal sample. The red, green, blue and orange bars below the patterns mark the Bragg peak positions for the common (hk00), ladder (hkl0), chain (hk0m) and satellite (hklm) Bragg peaks (from top to bottom), respectively.

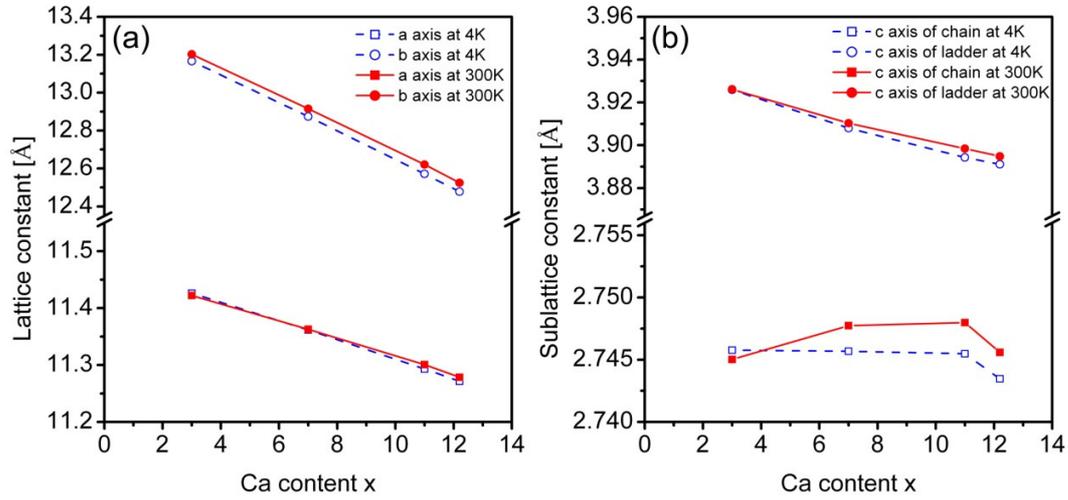

FIG. 3. The lattice parameters ($a$, $b$(a), $c_C$ and $c_L$(b)) of $Sr_{14-x}Ca_xCu_{24}O_{41}$ at 4K and 300K

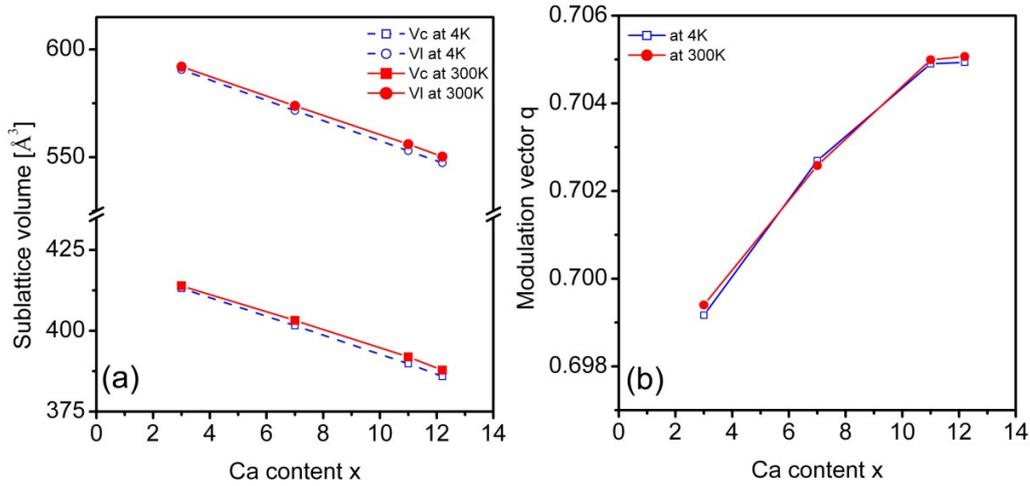

FIG. 4. The sublattice volumes (a) and modulated vectors (b) of $Sr_{14-x}Ca_xCu_{24}O_{41}$ at 4K and 300K

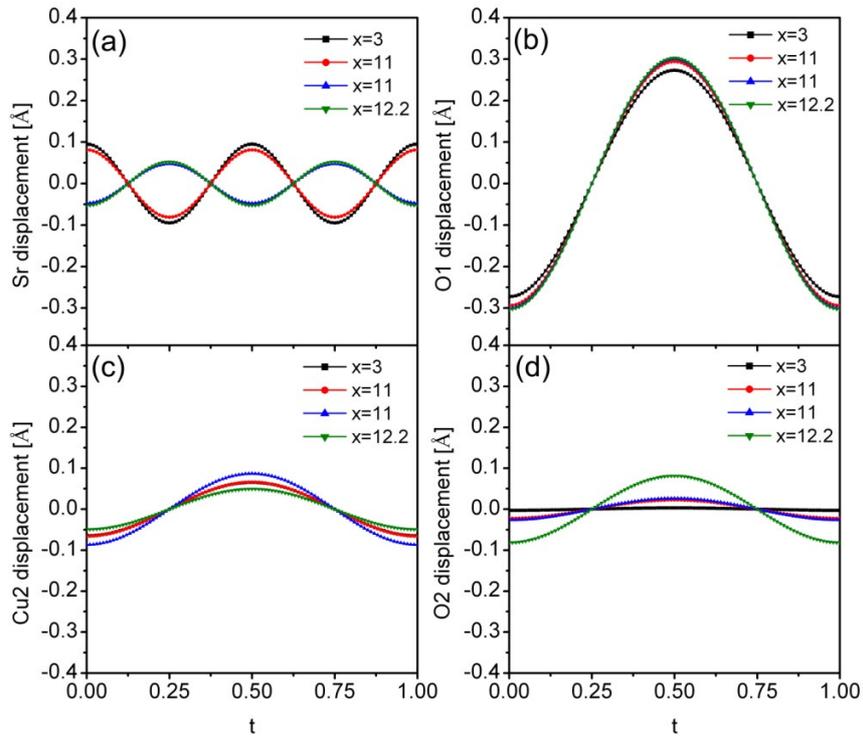

FIG. 5. Displacements of Sr(1)(a), O(1)(b), Cu(2)(c), O(2)(d) atoms modulated with $t$ along $b$ axis

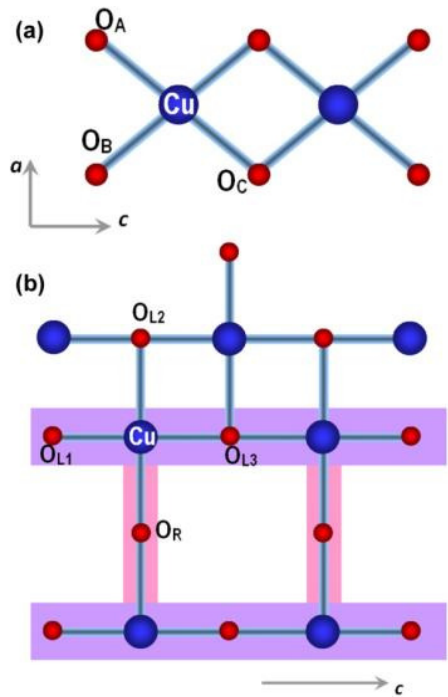

FIG. 6. The schematics of the two sublattices, chains (a) and ladders (b), in $Sr_{14-x}Ca_xCu_{24}O_{41}$. In (a), $O_A$, $O_B$ and $O_C$ denote the oxygen atoms on chains. In (b), $O_R$ denotes oxygen on the rung of the ladder, $O_{L1}$ and $O_{L3}$ denotes oxygen on the same leg of the ladder, and $O_{L2}$ denotes oxygen on the neighbor leg.

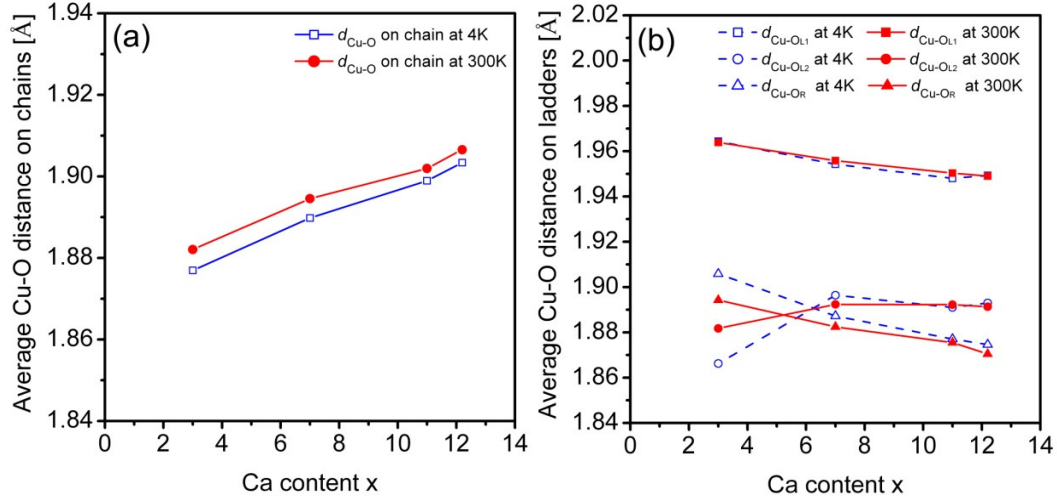

FIG. 7. The average Cu-O bond lengths on the chains (a) and ladders (b) of $Sr_{14-x}Ca_xCu_{24}O_{41}$

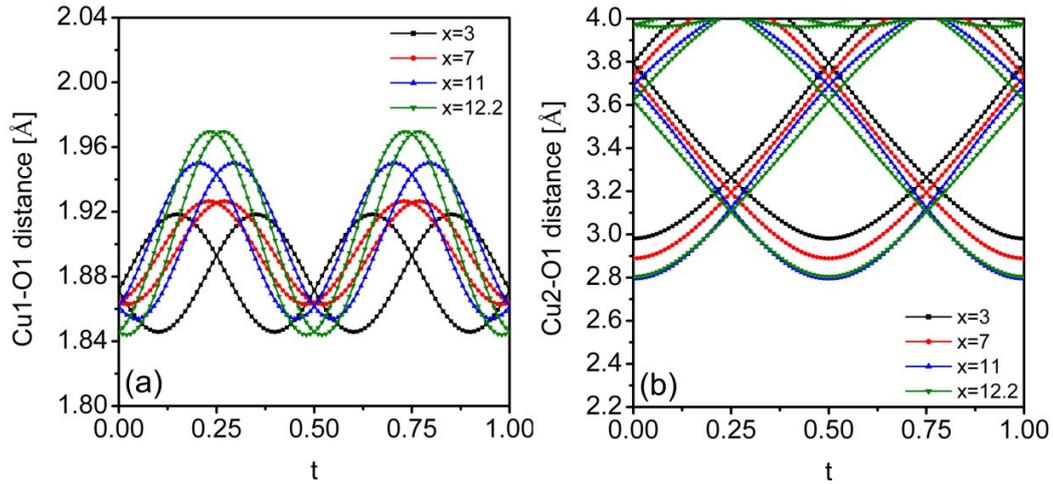

FIG. 8. Cu(1)-O(1) distances on the chains (a) and Cu(2)-O(1) distance (b) of $Sr_{14-x}Ca_xCu_{24}O_{41}$

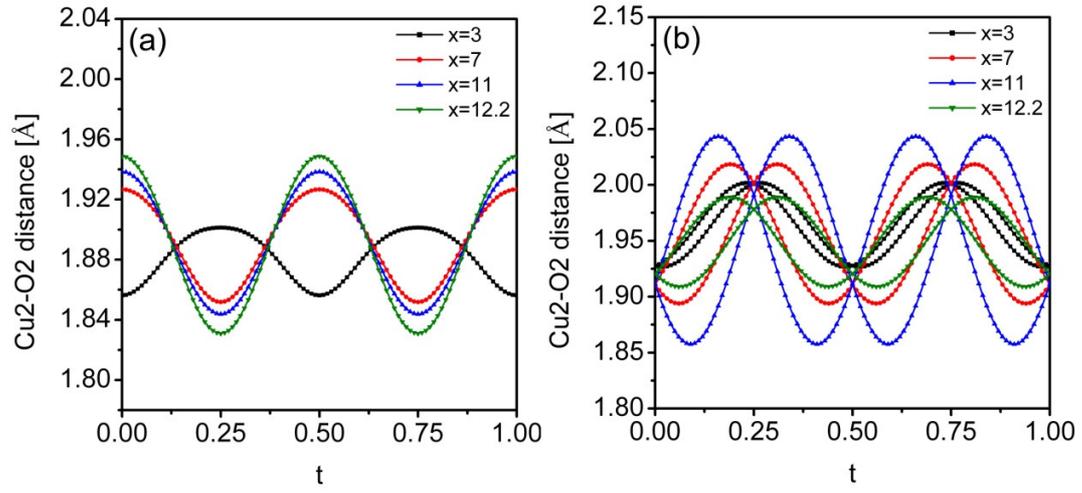

FIG. 9. Cu(2)-O(2) distance (i.e. Cu-O$_{L2}$ in FIG. 6 ) between legs (a) and Cu(2)-O(1) distance (i.e. Cu-O$_{L1}$/O$_{L3}$ in FIG. 6) along legs (b) in Sr$_{14-x}$Ca$_x$Cu$_{24}$O$_{41}$, modulating with *t*.

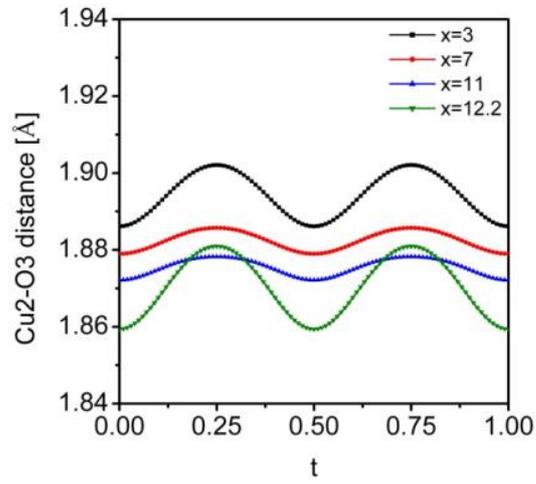

FIG. 10. Cu(2)-O(3) distance (i.e. Cu-O$_R$ in FIG. 6) on rungs of the ladders in Sr$_{14-x}$Ca$_x$Cu$_{24}$O$_{41}$

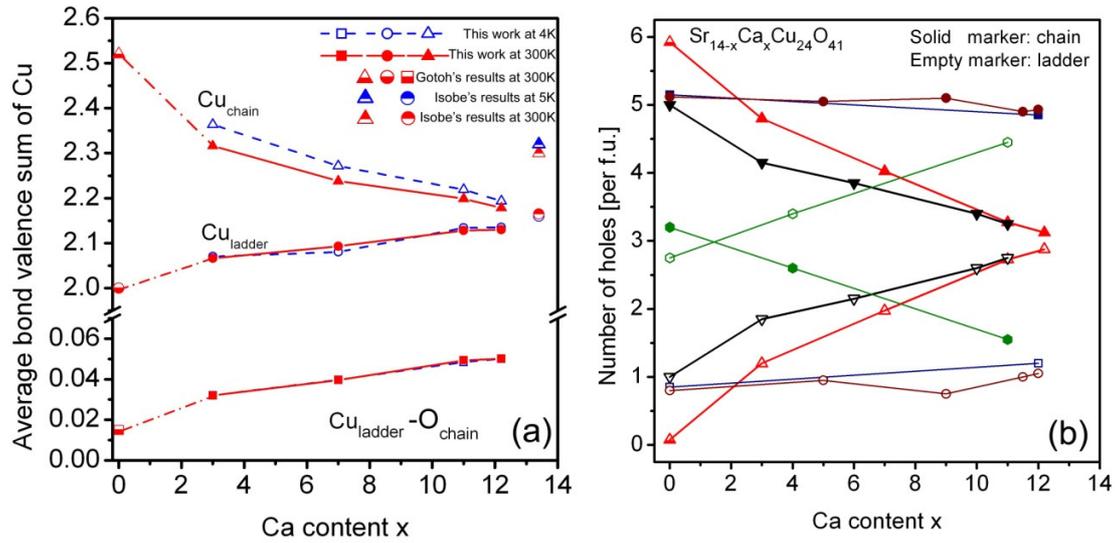

FIG. 11. (a) The average bond valence sum of Cu on the chains and ladders and bond valence of Cu(2)-O(1) with different Ca contents. The data with x=0 and 13.6 are from Ref [13] and Ref[25], respectively. (b) Hole number per formula in $Sr_{14-x}Ca_xCu_{24}O_{41}$ with different Ca content, which are reported by different authors using different methods. This graph is based on the comprehensive review of Vuletic *et al.*[2] with additional recently reported results. (black reverse triangle: Osafune's data[6]; brown circle: Nucker's data[30]; dark blue square: Piskunov's data[28]; green hexagon: Rusydi's data[29]; red triangle: this work. The data in this work has been normalized by assuming 6 holes per formula unit.)

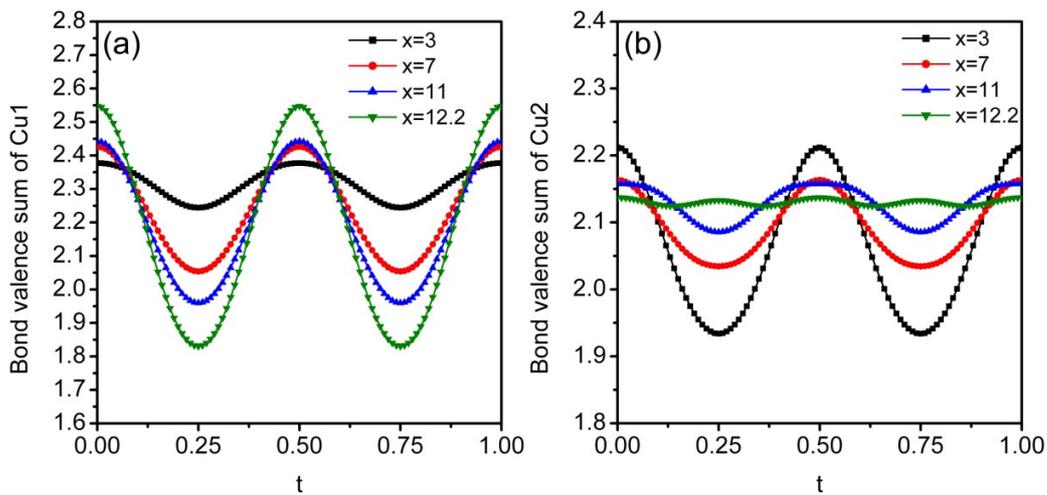

FIG. 12. The Cu-O bond valence sum modulating with *t* on the chains (a) and ladders (b) of Sr$_{14-x}$Ca$_x$Cu$_{24}$O$_{41}$

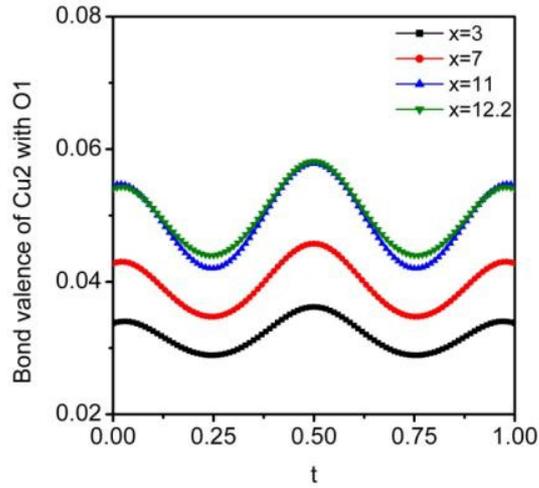

FIG. 13. The bond valence of Cu(2) and O(1) modulating with *t* in of Sr$_{14-x}$Ca$_x$Cu$_{24}$O$_{41}$

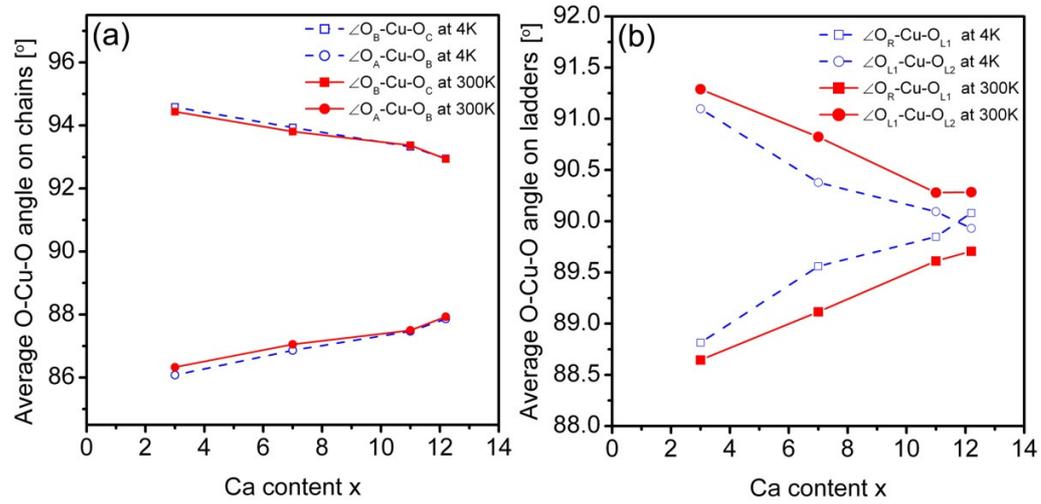

FIG. 14. The average O-Cu-O bond angles on the chains (a) and on the ladders (b) of Sr$_{14-x}$Ca$_x$Cu$_{24}$O$_{41}$

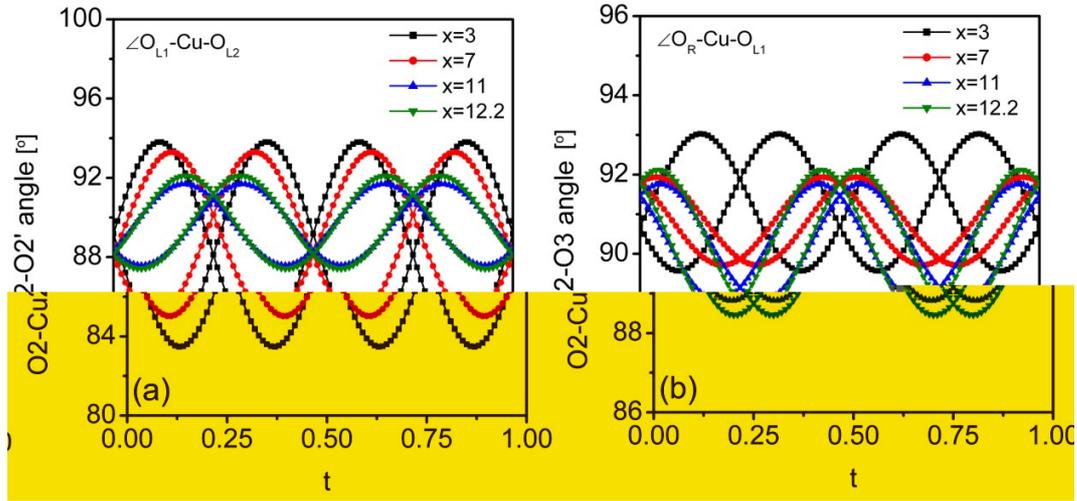

FIG. 15. The angles on the leg (a) and between leg and rung (b) modulating with $t$ on the ladders of $Sr_{14-x}Ca_xCu_{24}O_{41}$

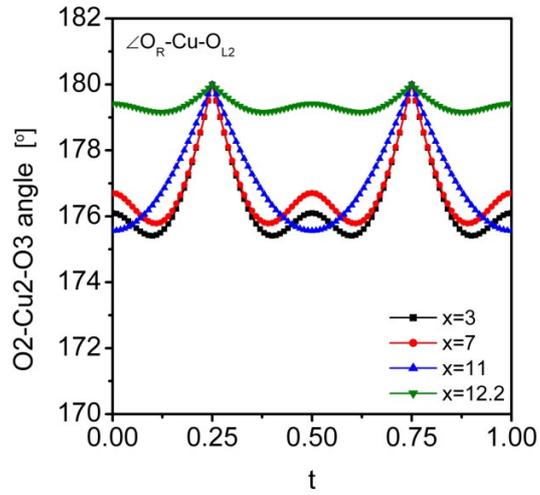

FIG. 16. The O-Cu-O angles along rungs modulating with $t$ on the ladders of $Sr_{14-x}Ca_xCu_{24}O_{41}$